\documentclass[aps,prd,11pt,twoside,tightenlines,nofootinbib,showpacs,preprint,superscriptaddress]{revtex4-1}
\usepackage[colorlinks=true]{hyperref}
\usepackage{xcolor}
\hypersetup{colorlinks=true, citecolor=blue, urlcolor=blue, linkcolor=blue}
\usepackage{amsmath,amssymb}
\usepackage[final]{graphicx}
\usepackage{subcaption}
\usepackage{slashed}
\usepackage{float}
\usepackage[toc,page]{appendix}

\usepackage{multirow}

\begin{document}

\arraycolsep1.5pt
\newcommand{\Ima}{\textrm{Im}}
\newcommand{\Rea}{\textrm{Re}}
\newcommand{\mev}{\textrm{ MeV}}
\newcommand{\gev}{\textrm{ GeV}}

\title{The hidden strange $B_{c}$-like molecular states}

\author{Zhong-Yu Wang}
\email{zhongyuwang@foxmail.com}
\affiliation{School of Physical Science and Technology, Lanzhou University, Lanzhou 730000, China}
\affiliation{Lanzhou Center for Theoretical Physics, Key Laboratory of Theoretical Physics of Gansu Province, and Key Laboratory of Quantum Theory and Applications of MoE, Lanzhou University, Lanzhou, Gansu 730000, China}

\author{Zhi-Feng Sun}
\email{sunzf@lzu.edu.cn}
\affiliation{School of Physical Science and Technology, Lanzhou University, Lanzhou 730000, China}
\affiliation{Lanzhou Center for Theoretical Physics, Key Laboratory of Theoretical Physics of Gansu Province, and Key Laboratory of Quantum Theory and Applications of MoE, Lanzhou University, Lanzhou, Gansu 730000, China}
\affiliation{Research Center for Hadron and CSR Physics, Lanzhou University and Institute of Modern Physics of CAS, Lanzhou 730000, China}
\affiliation{Frontiers Science Center for Rare Isotopes, Lanzhou University, Lanzhou, Gansu 730000, China}

\date{\today}

\begin{abstract}

With the chiral unitary approach, we evaluate the hidden strange $B_{c}$-like molecular states of $b\bar{c}s\bar{s}$ systems $\bar{B}_{s}\bar{D}_{s}$, $\bar{B}_{s}^{*}\bar{D}_{s}$, $\bar{B}_{s}\bar{D}_{s}^{*}$, and $\bar{B}_{s}^{*}\bar{D}_{s}^{*}$ coupled to the non-strange channels. The $S$-wave scattering amplitudes are calculated based on the vector meson exchange, four pseudoscalar mesons contact interactions, and four vector mesons contact interactions obtained from the extended local hidden gauge approach. We find six states below the threshold of the most relevant channel. The binding energies of these states are around $1-10$ MeV and the widths are around $0.2-0.7$ MeV. Our research is a supplement to the mass spectra of $B_{c}$-like states, which may be useful for the experimental search in the future.

\end{abstract}

\maketitle

\section{Introduction}

Searching for the $B_{c}$(-like) states is one of the important targets of particle physics, which provides opportunities to understand the nonperturbative behavior of strong interaction.  
In theory, the traditional $B_{c}$ meson spectra were predicted by quark models \cite{Godfrey:1985xj,Kwong:1990am,Eichten:1994gt,Zeng:1994vj,Gupta:1995ps,Fulcher:1998ka,Ebert:2002pp,Ikhdair:2003ry,Godfrey:2004ya,Ikhdair:2004hg,Ikhdair:2004tj,Soni:2017wvy,Eichten:2019gig,Li:2019tbn,Ortega:2020uvc}, QCD sum rules \cite{Gershtein:1994dxw,Wang:2012kw}, effective field theories \cite{Brambilla:2000db,Penin:2004xi,Peset:2018ria,Peset:2018jkf}, lattice QCD \cite{Allison:2004be,Dowdall:2012ab,Mathur:2018epb}, and continuum functional methods for QCD \cite{Yin:2019bxe,Chang:2019wpt,Chen:2020ecu}.
However, currently, there are only two $B_{c}$ states, i.e., the $B_{c}(6275)$ and $B_{c}(6872)$ listed in the Particle Data Group (PDG) \cite{ParticleDataGroup:2022pth}.
Excitingly, both the CMS \cite{CMS:2019uhm} and LHCb \cite{LHCb:2019bem} Collaborations found the excited $B_{c}^{+}(2^{1}S_{0})$ and $B_{c}^{*+}(2^{3}S_{1})$ states in the $B_{c}^{+}\pi^{+}\pi^{-}$ invariant mass spectrum recently.

These observations in experiments lead us to further explore the $B_{c}$-like states in theory.
Apart from the traditional mesons of $b\bar{c}$ quarks picture, the exotic $B_{c}$-like states were studied widely for a long time in the past. 
For instance, in view of the compact tetraquark states, Ref. \cite{Wu:2018xdi} studied the mass spectra of $B_{c}$-like states with $\bar{b}cq\bar{q}$, $\bar{b}cs\bar{q}$, $\bar{b}cq\bar{s}$, and $\bar{b}cs\bar{s}$ components based on the chromomagnetic interactions model. 
In Ref. \cite{Guo:2022crh}, with the improved chromomagnetic interactions model, the mass spectra of $b\bar{c}q\bar{q}$, $b\bar{c}s\bar{q}$, and $b\bar{c}s\bar{s}$ were predicted, particularly, the $S$-wave states with the quark content $b\bar{c}s\bar{s}$ and different quantum numbers were found around the thresholds of $B_{c}\phi$ and $\bar{B}_{s}^{(*)}\bar{D}_{s}^{(*)}$ channels.
Another important picture for studying $B_{c}$-like states is the hadronic molecular state generated by meson-meson interaction. 
Early in 2009, based on the QCD sum rule, the $\bar{B}\bar{D}$, $\bar{B}^{*}\bar{D}$, $\bar{B}\bar{D}^{*}$ and $\bar{B}^{*}\bar{D}^{*}$ molecular states mass spectra were predicted in Ref. \cite{Zhang:2009vs}. 
In 2012, the interactions of $B_{(s)}^{(*)}D_{(s)}^{(*)}$ systems were investigated with the one-boson-exchange model \cite{Sun:2012sy}, where it was found that the $\bar{b}cs\bar{s}$ bound states may be formed. 
Reference \cite{Sakai:2017avl} evaluated the $S$-wave interactions of the pseudoscalar and vector mesons systems with the quark contents $\bar{b}cq\bar{q}$ and $\bar{b}\bar{c}qq$ using the chiral unitary approach. 
Recently, the $bc\bar{s}\bar{q}$, $b\bar{c}s\bar{q}$, and $b\bar{c}\bar{s}q$ systems were investigated \cite{Liu:2023hrz}, where six $bc\bar{s}\bar{q}$ bound states were predicted, while no $b\bar{c}s\bar{q}$ or $b\bar{c}\bar{s}q$ states were found.
More discussions about the molecular states can be referred to the reviews of Refs. \cite{Chen:2016qju,Guo:2017jvc}.

With this background, we will utilize the chiral unitary approach to predict the mass spectra of hidden strange $B_{c}$-like states in the present work. 
The chiral unitary approach is successfully applicated in studying states generated by meson-meson and meson-baryon interactions, see, for instance, Refs. \cite{Oller:1997ti, Oset:1997it, Oller:2000ma, Oller:2000fj, Hyodo:2008xr}. 
On the one hand, in this approach, one take into account the Lagrangians constructed by the global chiral symmetry as well as the hidden gauge symmetry \cite{Wu:2010jy, Wu:2010vk, Xiao:2019gjd, Gamermann:2006nm, Molina:2010tx, Dai:2022ulk}. 
On the other hand, one can solve the Bethe-Salpeter equation of the on-shell approximation to deal with the nonperturbative physics restoring two-body unitarity in coupled channels \cite{Oller:1998hw, Molina:2009ct, Dias:2014pva, Oset:2022xji, Marse-Valera:2022khy}. 
In this paper, focusing on the $b\bar{c}s\bar{s}$ sector, we will consider the hidden strange channels $\bar{B}_{s}\bar{D}_{s}$, $\bar{B}_{s}^{*}\bar{D}_{s}$, $\bar{B}_{s}\bar{D}_{s}^{*}$, and $\bar{B}_{s}^{*}\bar{D}_{s}^{*}$ as well as the $\bar{B}\bar{D}$, $\bar{B}^{*}\bar{D}$, $\bar{B}\bar{D}^{*}$, and $\bar{B}^{*}\bar{D}^{*}$ channels. 
The interactions of light-vector-meson exchange, four pseudoscalar contact terms, and four vector mesons contact terms will be considered in these systems.
And we can see whether or not the hidden strange $B_{c}$-like molecular states exist by searching for the poles of the modulus square amplitudes in the complex energy plane, which correspond to the dynamically generated states.

This work is organized as follows. We will introduce the pseudoscalar-pseudoscalar, pseudoscalar-vector, and vector-vector mesons interactions formulas in Sec. \ref{sec:Formalism}. Next, the numerical results of scattering amplitudes and poles are shown in Sec. \ref{sec:Results}. Finally, we end this work with a brief summary in Sec. \ref{sec:Conclusions}.

\section{Formalism}
\label{sec:Formalism}

\subsection{Lagrangians}

In the present work, we study the $S$-wave interactions in the $\bar{B}_{s}\bar{D}_{s}$, $\bar{B}_{s}^{*}\bar{D}_{s}$, $\bar{B}_{s}\bar{D}_{s}^{*}$, and $\bar{B}_{s}^{*}\bar{D}_{s}^{*}$ hidden strange channels as well as another four channels $\bar{B}\bar{D}$, $\bar{B}^{*}\bar{D}$, $\bar{B}\bar{D}^{*}$, and $\bar{B}^{*}\bar{D}^{*}$. The vertices $VPP$ and $VVV$ are needed to calculate the light-vector-meson exchange potentials. Moreover, the contact terms $PPPP$ and $VVVV$ are also needed. Here, $V$ and $P$ denote the vector and pseudoscalar mesons fields, respectively. The relevant local hidden gauge Lagrangians are constructed by adopting the hidden gauge symmetry and chiral symmetry, which are shown in the following \cite{Bando:1984ej, Bando:1987br, Sun:2018zqs}
\begin{equation}
	\begin{aligned} 
		\mathcal{L}_{V P P}=-i g\left\langle\left[P, \partial_\mu P\right] V^\mu\right\rangle,
	\end{aligned}
	\label{eq:LVVP}
\end{equation}
\begin{equation}
	\begin{aligned} 
		\mathcal{L}_{V V V}=i g\left\langle\left(V^\mu \partial_\nu V_\mu-\partial_\nu V^\mu V_\mu\right) V^\nu\right\rangle,
	\end{aligned}
	\label{eq:LVVV}
\end{equation}
\begin{equation}
	\begin{aligned} 
		\mathcal{L}_{P P P P}=-\frac{1}{24f_{\pi}^2}\left\langle\left[P, \partial_\mu P\right]\left[P, \partial^\mu P\right] \right\rangle,
	\end{aligned}
	\label{eq:LPPPP}
\end{equation}
\begin{equation}
	\begin{aligned} 
		\mathcal{L}_{V V V V}=\frac{g^2}{2}\left\langle V_\mu V_\nu V^\mu V^\nu-V_\nu V_\mu V^\mu V^\nu\right\rangle.
	\end{aligned}
	\label{eq:LVVVV}
\end{equation}
The coupling constant $g=M_{V}/(2f_{\pi})$ with $M_{V}$ the mass of the vector meson taken as $800$ MeV \cite{Liu:2023hrz, Wang:2022aga} and $f_{\pi}=93$ MeV the decay constant of pion. The symbol $\left\langle...\right\rangle$ means the trace of the matrices. In this work, we consider the $SU(5)$ flavor symmetry since we study the interactions between the charmed and bottomed mesons. The fields $P$ and $V^{\mu}$ in Eqs. (\ref{eq:LVVP})-(\ref{eq:LVVVV}) are written as
\begin{equation}
	P=\left(\begin{array}{ccccc}
		\frac{\eta}{\sqrt{3}}+\frac{\eta^{\prime}}{\sqrt{6}}+\frac{\pi^0}{\sqrt{2}} & \pi^{+} & K^{+} & \bar{D}^0 & B^{+} \\
		\pi^{-} & \frac{\eta}{\sqrt{3}}+\frac{\eta^{\prime}}{\sqrt{6}}-\frac{\pi^0}{\sqrt{2}} & K^0 & D^{-} & B^0 \\
		K^{-} & \bar{K}^0 & -\frac{\eta}{\sqrt{3}}+\sqrt{\frac{2}{3}}\eta^{\prime} & D_s^{-} & B_s^0 \\
		D^0 & D^{+} & D_s^{+} & \eta_c & B_c^{+} \\
		B^{-} & \bar{B}^0 & \bar{B}_s^0 & B_c^{-} & \eta_b
	\end{array}\right),
\end{equation}
\begin{equation}
	V^{\mu}=\left(\begin{array}{ccccc}
		\frac{\omega+\rho^0}{\sqrt{2}} & \rho^{+} & K^{*+} & \bar{D}^{* 0} & B^{*+} \\
		\rho^{-} & \frac{\omega-\rho^0}{\sqrt{2}} & K^{* 0} & D^{*-} & B^{* 0} \\
		K^{*-} & \bar{K}^{* 0} & \phi & D_s^{*-} & B_s^{* 0} \\
		D^{* 0} & D^{*+} & D_s^{*+} & J / \psi & B_c^{*+} \\
		B^{*-} & \bar{B}^{* 0} & \bar{B}_s^{* 0} & B_c^{*-} & \Upsilon
	\end{array}\right)^{\mu},
\end{equation}
respectively.

In the following, we will use the above Lagrangians and fields to calculate the potentials of $\bar{B}^{(*)}\bar{D}^{(*)}$/$\bar{B}_{s}^{(*)}\bar{D}_{s}^{(*)}$ systems with the isospin $I=0$, where we take the isospin phase convention $|B^{(*)-}\rangle=-| 1/2,-1/2\rangle$, $|\bar{B}^{(*)0}\rangle=| 1/2,1/2\rangle$, $|D^{(*)-}\rangle=| 1/2,-1/2\rangle$, $|\bar{D}^{(*)0}\rangle=| 1/2,1/2\rangle$, $|\bar{B}^{(*)0}_{s}\rangle=|0,0\rangle$, and $|D^{(*)-}_{s}\rangle=|0,0\rangle$. 

\subsection{The $\bar{B}\bar{D}$/$\bar{B}_{s}\bar{D}_{s}$ system}

\begin{figure}[htbp]
	\begin{subfigure}{0.45\textwidth}
		\centering
		\includegraphics[width=1\linewidth,trim=150 580 250 120,clip]{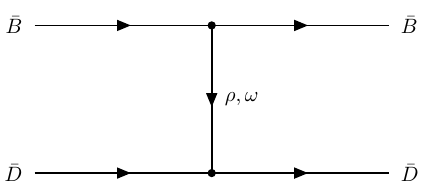} 
		\caption{\footnotesize}
		\label{fig:PP11t}
	\end{subfigure}
	\quad
	\quad
	\begin{subfigure}{0.45\textwidth}  
		\centering 
		\includegraphics[width=1\linewidth,trim=150 580 250 120,clip]{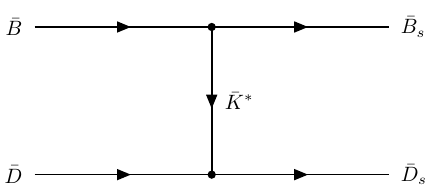} 
		\caption{\footnotesize}
		\label{fig:PP12t}  
	\end{subfigure}	
	\begin{subfigure}{0.45\textwidth}  
		\centering 
		\includegraphics[width=1\linewidth,trim=150 580 250 120,clip]{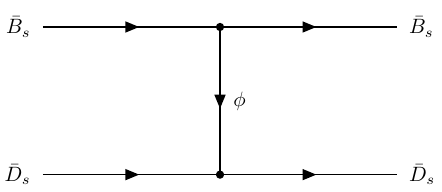} 
		\caption{\footnotesize}
		\label{fig:PP22t}  
	\end{subfigure}	
	\caption{The light-vector-meson exchange contribution for the two pseudoscalar mesons coupled system.}
	\label{fig:VPP}
\end{figure}

In this subsection, we consider the $\bar{B}\bar{D}$ and $\bar{B}_{s}\bar{D}_{s}$ coupled system. The elementary interactions can be obtained from the t-channel contributions, whose corresponding diagrams are shown in Fig. \ref{fig:VPP}. Note that we neglect the heavy-vector-meson exchange processes, since they are too massive and the amplitudes are suppressed. Using the Lagrangians in Eq. (\ref{eq:LVVP}), we obtain the effective potentials of the two pseudoscalar mesons by the light-vector-meson exchange as follows

\begin{equation}
	\begin{aligned} 
		v^{ex}_{\bar{B}\bar{D}\rightarrow\bar{B}\bar{D}}=&
		\left(-\frac{3}{2m^{2}_{\rho}}-\frac{1}{2m^{2}_{\omega}}\right)g^2(k_{1}+k_{3})\cdot(k_{2}+k_{4}), \\
		v^{ex}_{\bar{B}\bar{D}\rightarrow\bar{B}_{s}\bar{D}_{s}}=&
		-\frac{\sqrt{2}}{m^{2}_{\bar{K}^{*}}+m^{2}_{\bar{B}_{s}}-m^{2}_{\bar{B}}}g^2(k_{1}+k_{3})\cdot(k_{2}+k_{4}), \\
		v^{ex}_{\bar{B}_{s}\bar{D}_{s}\rightarrow\bar{B}_{s}\bar{D}_{s}}=&
		-\frac{1}{m^{2}_{\phi}}g^2(k_{1}+k_{3})\cdot(k_{2}+k_{4}), 
	\end{aligned}
	\label{eq:VPP}
\end{equation}
where $k_{1}$ and $k_{2}$ are the four-momentums of the initial mesons, and $k_{3}$ and $k_{4}$ are the four-momentums of the final mesons. As done in Refs. \cite{Molina:2008jw, Geng:2008gx}, in $v_{\bar{B}\bar{D}\rightarrow\bar{B}\bar{D}}$ and $v_{\bar{B}_{s}\bar{D}_{s}\rightarrow\bar{B}_{s}\bar{D}_{s}}$, we take
\begin{equation}
	\begin{aligned} 
		\frac{1}{q^2-m_V^2+i \epsilon} \approx-\frac{1}{m_V^2},
	\end{aligned}
	\label{eq:diagonal}
\end{equation}
where the transferred four-momentum $q^{\mu}$ in the propagator is neglected, since it is very small when we only focus on the energy region near the threshold of $\bar{B}_{s}\bar{D}_{s}$.
However, this reduction does not work well for $v_{\bar{B}\bar{D}\rightarrow\bar{B}_{s}\bar{D}_{s}}$ due to the large mass difference between the initial and final mesons. And we adopt the following approximation \cite{Bayar:2022dqa}
\begin{equation}
	\begin{aligned} 
		q^2=(p_{\bar{B}}-p_{\bar{B}_{s}})^2=p^2_{\bar{B}}+p^2_{\bar{B}_{s}}-2p_{\bar{B}}\cdot p_{\bar{B}_{s}}=m^2_{\bar{B}}-m^2_{\bar{B}_{s}}.
	\end{aligned}
	\label{eq:q0}
\end{equation}
In this way, we have
\begin{equation}
	\begin{aligned} 
		\frac{1}{q^2-m^{2}_{\bar{K}^{*}}+i \epsilon} \approx-\frac{1}{m^{2}_{\bar{K}^{*}}+m^{2}_{\bar{B}_{s}}-m^{2}_{\bar{B}}}.
	\end{aligned}
	\label{eq:offdiagonal}
\end{equation}
Similar approximations are used in the pseudoscalar-vector and vector-vector systems as well. After performing the partial wave projection, one can obtain the $S$-wave potentials as functions of the rest frame energy.

\begin{figure}[htbp]
	\begin{subfigure}{0.45\textwidth}
		\centering
		\includegraphics[width=1\linewidth,trim=150 580 250 120,clip]{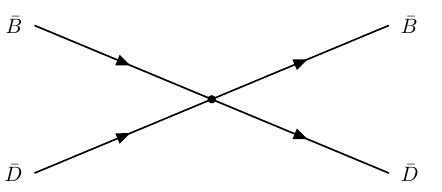} 
		\caption{\footnotesize}
		\label{fig:PPPP11}
	\end{subfigure}
	\quad
	\quad
	\begin{subfigure}{0.45\textwidth}  
		\centering 
		\includegraphics[width=1\linewidth,trim=150 580 250 120,clip]{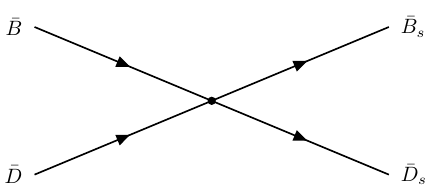} 
		\caption{\footnotesize}
		\label{fig:PPPP12}  
	\end{subfigure}	
	\begin{subfigure}{0.45\textwidth}  
		\centering 
		\includegraphics[width=1\linewidth,trim=150 580 250 120,clip]{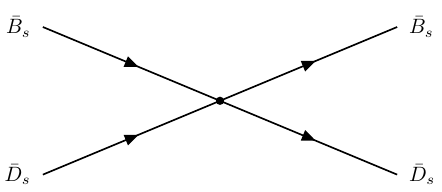} 
		\caption{\footnotesize}
		\label{fig:PPPP22}  
	\end{subfigure}	
	\caption{The contact term connecting four pseudoscalar mesons.}
	\label{fig:PPPP}
\end{figure}

In addition, the interactions contributed by four pseudoscalar meson contact terms are shown in Fig. \ref{fig:PPPP}. The corresponding potentials are as follows

\begin{equation}
	\begin{aligned} 
		v^{co}_{\bar{B}\bar{D}\rightarrow\bar{B}\bar{D}}=&
		-\frac{1}{6f_{\pi}^{2}}\left(-k_1\cdot k_2-2k_1\cdot k_4+k_1\cdot k_3+k_2\cdot k_4-2k_2\cdot k_3-k_3\cdot k_4\right), \\
		v^{co}_{\bar{B}\bar{D}\rightarrow\bar{B}_{s}\bar{D}_{s}}=&
		-\frac{\sqrt{2}}{12f_{\pi}^{2}}\left(-k_1\cdot k_2-2k_1\cdot k_4+k_1\cdot k_3+k_2\cdot k_4-2k_2\cdot k_3-k_3\cdot k_4\right), \\
		v^{co}_{\bar{B}_{s}\bar{D}_{s}\rightarrow\bar{B}_{s}\bar{D}_{s}}=&
		-\frac{1}{12f_{\pi}^{2}}\left(-k_1\cdot k_2-2k_1\cdot k_4+k_1\cdot k_3+k_2\cdot k_4-2k_2\cdot k_3-k_3\cdot k_4\right). 
	\end{aligned}
	\label{eq:PPPP}
\end{equation}

\subsection{The $\bar{B}^{*}\bar{D}$/$\bar{B}\bar{D}^{*}$/$\bar{B}_{s}^{*}\bar{D}_{s}$/$\bar{B}_{s}\bar{D}_{s}^{*}$ system}

\begin{figure}[htbp]
	\begin{subfigure}{0.45\textwidth}
		\centering
		\includegraphics[width=1\linewidth,trim=150 580 250 120,clip]{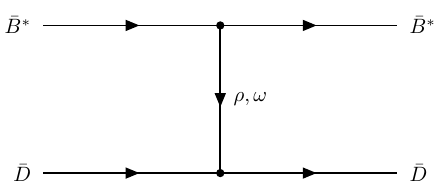} 
		\caption{\footnotesize}
		\label{fig:VP11t}
	\end{subfigure}
	\quad
	\quad
	\begin{subfigure}{0.45\textwidth}  
		\centering 
		\includegraphics[width=1\linewidth,trim=150 580 250 120,clip]{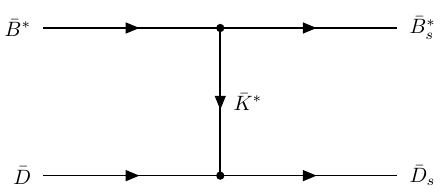} 
		\caption{\footnotesize}
		\label{fig:VP13t}  
	\end{subfigure}	
	\begin{subfigure}{0.45\textwidth}
		\centering
		\includegraphics[width=1\linewidth,trim=150 580 250 120,clip]{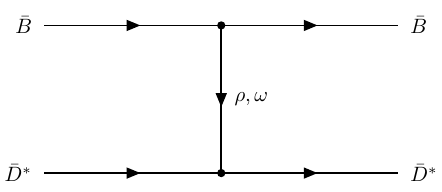} 
		\caption{\footnotesize}
		\label{fig:VP22t}
	\end{subfigure}
	\quad
	\quad
	\begin{subfigure}{0.45\textwidth}  
		\centering 
		\includegraphics[width=1\linewidth,trim=150 580 250 120,clip]{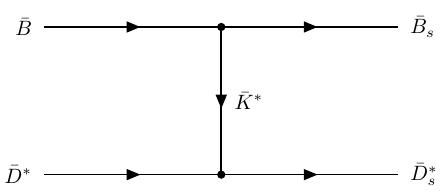} 
		\caption{\footnotesize}
		\label{fig:VP24t}  
	\end{subfigure}
	\begin{subfigure}{0.45\textwidth}
		\centering
		\includegraphics[width=1\linewidth,trim=150 580 250 120,clip]{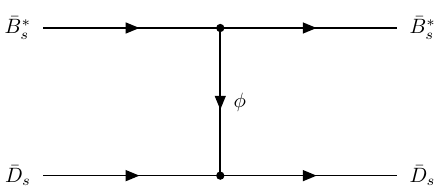} 
		\caption{\footnotesize}
		\label{fig:VP33t}
	\end{subfigure}
	\quad
	\quad
	\begin{subfigure}{0.45\textwidth}  
		\centering 
		\includegraphics[width=1\linewidth,trim=150 580 250 120,clip]{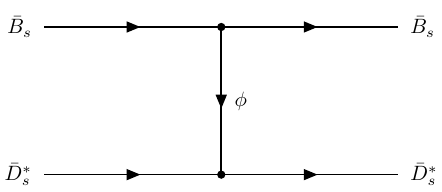} 
		\caption{\footnotesize}
		\label{fig:VP44t}  
	\end{subfigure}
	\caption{The light-vector-meson exchange contribution for the pseudoscalar and vector mesons coupled system.}
	\label{fig:VVP}
\end{figure}

In this subsection, we take into account the interactions between the pseudoscalar and vector mesons, where there exist four coupled channels, i.e., $\bar{B}^{*}\bar{D}$, $\bar{B}\bar{D}^{*}$, $\bar{B}_{s}^{*}\bar{D}_{s}$, and $\bar{B}_{s}\bar{D}_{s}^{*}$. The quantum number of this system is $I(J^{P})=0(1^{+})$. The diagrams of the possible processes are shown in Fig. \ref{fig:VVP}. Note that there is no vector meson exchange interaction between $\bar{B}^{*}\bar{D}$/$\bar{B}_{s}^{*}\bar{D}_{s}$ and $\bar{B}\bar{D}^{*}$/$\bar{B}_{s}\bar{D}_{s}^{*}$, because the vertex $VVP$ is anomalous. Based on Eqs. (\ref{eq:LVVP}) and (\ref{eq:LVVV}), we get the effective potentials with $I=0$
\begin{equation}
	\begin{aligned} 
		v_{\bar{B}^{*}\bar{D}\rightarrow\bar{B}^{*}\bar{D}}=&
		\left(-\frac{3}{2m^{2}_{\rho}}-\frac{1}{2m^{2}_{\omega}}\right)g^2(k_{1}+k_{3})\cdot(k_{2}+k_{4})(\vec{\epsilon}_{1}\cdot\vec{\epsilon}_{3}^{\,\dagger}), \\
		v_{\bar{B}^{*}\bar{D}\rightarrow\bar{B}\bar{D}^{*}}=&0, \\
		v_{\bar{B}^{*}\bar{D}\rightarrow\bar{B}_{s}^{*}\bar{D}_{s}}=&
		-\frac{\sqrt{2}}{m^{2}_{\bar{K}^{*}}+m^{2}_{\bar{B}_{s}^{*}}-m^{2}_{\bar{B}^{*}}}g^2(k_{1}+k_{3})\cdot(k_{2}+k_{4})(\vec{\epsilon}_{1}\cdot\vec{\epsilon}_{3}^{\,\dagger}), \\
		v_{\bar{B}^{*}\bar{D}\rightarrow\bar{B}_{s}\bar{D}_{s}^{*}}=&0, \\
		v_{\bar{B}\bar{D}^{*}\rightarrow\bar{B}\bar{D}^{*}}=&		\left(-\frac{3}{2m^{2}_{\rho}}-\frac{1}{2m^{2}_{\omega}}\right)g^2(k_{1}+k_{3})\cdot(k_{2}+k_{4})(\vec{\epsilon}_{2}\cdot\vec{\epsilon}_{4}^{\,\dagger}), \\
		v_{\bar{B}\bar{D}^{*}\rightarrow\bar{B}_{s}^{*}\bar{D}_{s}}=&0, \\
		v_{\bar{B}\bar{D}^{*}\rightarrow\bar{B}_{s}\bar{D}_{s}^{*}}=&
		-\frac{\sqrt{2}}{m^{2}_{\bar{K}^{*}}+m^{2}_{\bar{B}_{s}}-m^{2}_{\bar{B}}}g^2(k_{1}+k_{3})\cdot(k_{2}+k_{4})(\vec{\epsilon}_{2}\cdot\vec{\epsilon}_{4}^{\,\dagger}), \\
		v_{\bar{B}_{s}^{*}\bar{D}_{s}\rightarrow\bar{B}_{s}^{*}\bar{D}_{s}}=&
		-\frac{1}{m^{2}_{\phi}}g^2(k_{1}+k_{3})\cdot(k_{2}+k_{4})(\vec{\epsilon}_{1}\cdot\vec{\epsilon}_{3}^{\,\dagger}), \\
		v_{\bar{B}_{s}^{*}\bar{D}_{s}\rightarrow\bar{B}_{s}\bar{D}_{s}^{*}}=&0, \\
		v_{\bar{B}_{s}\bar{D}_{s}^{*}\rightarrow\bar{B}_{s}\bar{D}_{s}^{*}}=&
		-\frac{1}{m^{2}_{\phi}}g^2(k_{1}+k_{3})\cdot(k_{2}+k_{4})(\vec{\epsilon}_{2}\cdot\vec{\epsilon}_{4}^{\,\dagger}). \\
	\end{aligned}
	\label{eq:VVP}
\end{equation}
In the above equations, $\vec{\epsilon}_{1}\cdot\vec{\epsilon}_{3}^{\,\dagger}$ and $\vec{\epsilon}_{2}\cdot\vec{\epsilon}_{4}^{\,\dagger}$ are the products of the polarization vectors of the initial and final mesons. Since the three-momentums of the external mesons are small, we take $\epsilon^{0}=0$.

\subsection{The $\bar{B}^{*}\bar{D}^{*}$/$\bar{B}_{s}^{*}\bar{D}_{s}^{*}$ system}

There are also only two channels $\bar{B}^{*}\bar{D}^{*}$ and $\bar{B}_{s}^{*}\bar{D}_{s}^{*}$ for this system, whose quantum numbers could be $I(J^{P})=0(0^{+})$, $0(1^{+})$, and $0(2^{+})$. 

\begin{figure}[htbp]
	\begin{subfigure}{0.45\textwidth}
		\centering
		\includegraphics[width=1\linewidth,trim=150 580 250 120,clip]{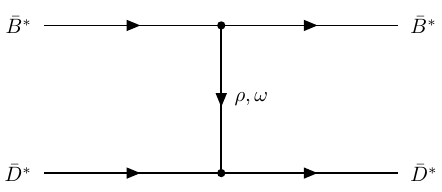} 
		\caption{\footnotesize}
		\label{fig:VV11t}
	\end{subfigure}
	\quad
	\quad
	\begin{subfigure}{0.45\textwidth}  
		\centering 
		\includegraphics[width=1\linewidth,trim=150 580 250 120,clip]{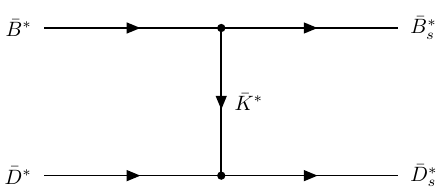} 
		\caption{\footnotesize}
		\label{fig:VV12t}  
	\end{subfigure}	
	\begin{subfigure}{0.45\textwidth}  
		\centering 
		\includegraphics[width=1\linewidth,trim=150 580 250 120,clip]{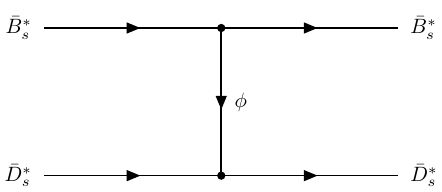} 
		\caption{\footnotesize}
		\label{fig:VV22t}  
	\end{subfigure}	
	\caption{The light-vector-meson exchange contribution for the two vector mesons coupled system.}
	\label{fig:VVV}
\end{figure}

As shown in Fig. \ref{fig:VVV}, the elementary interactions from the t-channel contributions are provided by the light-vector-meson exchange.
\begin{figure}[htbp]
	\begin{subfigure}{0.45\textwidth}
		\centering
		\includegraphics[width=1\linewidth,trim=150 580 250 120,clip]{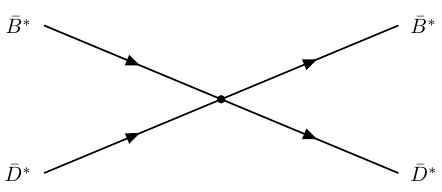} 
		\caption{\footnotesize}
		\label{fig:VVVV11}
	\end{subfigure}
	\quad
	\quad
	\begin{subfigure}{0.45\textwidth}  
		\centering 
		\includegraphics[width=1\linewidth,trim=150 580 250 120,clip]{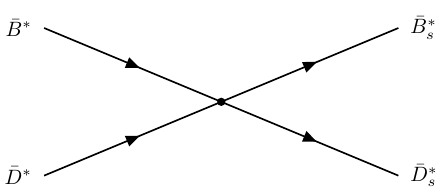} 
		\caption{\footnotesize}
		\label{fig:VVVV12}  
	\end{subfigure}	
	\begin{subfigure}{0.45\textwidth}  
		\centering 
		\includegraphics[width=1\linewidth,trim=150 580 250 120,clip]{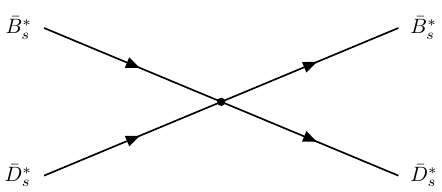} 
		\caption{\footnotesize}
		\label{fig:VVVV22}  
	\end{subfigure}	
	\caption{The contact term connecting four vector mesons.}
	\label{fig:VVVV}
\end{figure}

According to the Lagrangians in Eq. (\ref{eq:LVVV}), we obtain the effective potentials with $I=0$ as follows
\begin{equation}
	\begin{aligned} 
		v^{ex}_{\bar{B}^{*}\bar{D}^{*}\rightarrow\bar{B}^{*}\bar{D}^{*}}=&
		\left(-\frac{3}{2m^{2}_{\rho}}-\frac{1}{2m^{2}_{\omega}}\right)g^2(k_{1}+k_{3})\cdot(k_{2}+k_{4})(\vec{\epsilon}_{1}\cdot\vec{\epsilon}_{3}^{\,\dagger})(\vec{\epsilon}_{2}\cdot\vec{\epsilon}_{4}^{\,\dagger}), \\
		v^{ex}_{\bar{B}^{*}\bar{D}^{*}\rightarrow\bar{B}_{s}^{*}\bar{D}_{s}^{*}}=&
		-\frac{\sqrt{2}}{m^{2}_{\bar{K}^{*}}+m^{2}_{\bar{B}_{s}^{*}}-m^{2}_{\bar{B}^{*}}}g^2(k_{1}+k_{3})\cdot(k_{2}+k_{4})(\vec{\epsilon}_{1}\cdot\vec{\epsilon}_{3}^{\,\dagger})(\vec{\epsilon}_{2}\cdot\vec{\epsilon}_{4}^{\,\dagger}), \\
		v^{ex}_{\bar{B}_{s}^{*}\bar{D}_{s}^{*}\rightarrow\bar{B}_{s}^{*}\bar{D}_{s}^{*}}=&
		-\frac{1}{m^{2}_{\phi}}g^2(k_{1}+k_{3})\cdot(k_{2}+k_{4})(\vec{\epsilon}_{1}\cdot\vec{\epsilon}_{3}^{\,\dagger})(\vec{\epsilon}_{2}\cdot\vec{\epsilon}_{4}^{\,\dagger}).
	\end{aligned}
	\label{eq:VVV}
\end{equation}
Additionally, the interactions provided by the contact terms are shown in Fig. \ref{fig:VVVV}, and the corresponding potentials read
\begin{equation}
	\begin{aligned} 
		v^{co}_{\bar{B}^{*}\bar{D}^{*}\rightarrow\bar{B}^{*}\bar{D}^{*}}=&		-2g^2\left[-(\vec{\epsilon}_{1}\cdot\vec{\epsilon}_{2})(\vec{\epsilon}_{3}^{\,\dagger}\cdot\vec{\epsilon}_{4}^{\,\dagger})-(\vec{\epsilon}_{1}\cdot\vec{\epsilon}_{3}^{\,\dagger})(\vec{\epsilon}_{2}\cdot\vec{\epsilon}_{4}^{\,\dagger})+2(\vec{\epsilon}_{1}\cdot\vec{\epsilon}_{4}^{\,\dagger})(\vec{\epsilon}_{2}\cdot\vec{\epsilon}_{3}^{\,\dagger})\right], \\
		v^{co}_{\bar{B}^{*}\bar{D}^{*}\rightarrow\bar{B}_{s}^{*}\bar{D}_{s}^{*}}=&
		-\sqrt{2}g^2\left[-(\vec{\epsilon}_{1}\cdot\vec{\epsilon}_{2})(\vec{\epsilon}_{3}^{\,\dagger}\cdot\vec{\epsilon}_{4}^{\,\dagger})-(\vec{\epsilon}_{1}\cdot\vec{\epsilon}_{3}^{\,\dagger})(\vec{\epsilon}_{2}\cdot\vec{\epsilon}_{4}^{\,\dagger})+2(\vec{\epsilon}_{1}\cdot\vec{\epsilon}_{4}^{\,\dagger})(\vec{\epsilon}_{2}\cdot\vec{\epsilon}_{3}^{\,\dagger})\right], \\
		v^{co}_{\bar{B}_{s}^{*}\bar{D}_{s}^{*}\rightarrow\bar{B}_{s}^{*}\bar{D}_{s}^{*}}=&
		-g^2\left[-(\vec{\epsilon}_{1}\cdot\vec{\epsilon}_{2})(\vec{\epsilon}_{3}^{\,\dagger}\cdot\vec{\epsilon}_{4}^{\,\dagger})-(\vec{\epsilon}_{1}\cdot\vec{\epsilon}_{3}^{\,\dagger})(\vec{\epsilon}_{2}\cdot\vec{\epsilon}_{4}^{\,\dagger})+2(\vec{\epsilon}_{1}\cdot\vec{\epsilon}_{4}^{\,\dagger})(\vec{\epsilon}_{2}\cdot\vec{\epsilon}_{3}^{\,\dagger})\right].
	\end{aligned}
	\label{eq:VVVV}
\end{equation}
With the following spin projection operators \cite{Molina:2008jw}, one can obtain the potentials of spin $J=0$, $1$, and $2$,
\begin{equation}
	\begin{aligned}
		\mathcal{P}^{(0)}=& \frac{1}{3}(\vec{\epsilon}_{1}\cdot\vec{\epsilon}_{2})(\vec{\epsilon}_{3}^{\,\dagger}\cdot\vec{\epsilon}_{4}^{\,\dagger}), \\
		\mathcal{P}^{(1)}=& \frac{1}{2}\left[(\vec{\epsilon}_{1}\cdot\vec{\epsilon}_{3}^{\,\dagger})(\vec{\epsilon}_{2}\cdot\vec{\epsilon}_{4}^{\,\dagger})-(\vec{\epsilon}_{1}\cdot\vec{\epsilon}_{4}^{\,\dagger})(\vec{\epsilon}_{2}\cdot\vec{\epsilon}_{3}^{\,\dagger})\right], \\
		\mathcal{P}^{(2)}=& \frac{1}{2}\left[(\vec{\epsilon}_{1}\cdot\vec{\epsilon}_{3}^{\,\dagger})(\vec{\epsilon}_{2}\cdot\vec{\epsilon}_{4}^{\,\dagger})+(\vec{\epsilon}_{1}\cdot\vec{\epsilon}_{4}^{\,\dagger})(\vec{\epsilon}_{2}\cdot\vec{\epsilon}_{3}^{\,\dagger})\right]-\frac{1}{3}(\vec{\epsilon}_{1}\cdot\vec{\epsilon}_{2})(\vec{\epsilon}_{3}^{\,\dagger}\cdot\vec{\epsilon}_{4}^{\,\dagger}).
	\end{aligned}
\end{equation}

\subsection{The Bethe-Salpeter equation}

Once the effective potentials are obtained, we solve the Bethe-Salpeter equation with the on-shell approximation, which is shown below \cite{Oller:1997ti, Oset:1997it, Oller:2000ma},
\begin{equation}
	\begin{aligned} 
		T = [1-vG]^{-1}v. 
	\end{aligned}
	\label{eq:BSE}
\end{equation}
Here, $v$ denotes the effective potential matrix. $G$ is a diagonal matrix composed of the loop functions, for which the element with the dimensional regularization is \cite{Oller:1998zr, Alvarez-Ruso:2010rqm, Guo:2016zep}
\begin{equation}
	\begin{aligned}
		G_{ii}(s)=& \frac{1}{16 \pi^{2}}\left\{a_{ii}(\mu)+\ln \frac{m_{1}^{2}}{\mu^{2}}+\frac{m_{2}^{2}-m_{1}^{2}+s}{2 s} \ln \frac{m_{2}^{2}}{m_{1}^{2}}\right.\\
		&+\frac{q_{cmi}(s)}{\sqrt{s}}\left[\ln \left(s-\left(m_{2}^{2}-m_{1}^{2}\right)+2 q_{cmi}(s) \sqrt{s}\right)\right.\\
		&+\ln \left(s+\left(m_{2}^{2}-m_{1}^{2}\right)+2 q_{cmi}(s) \sqrt{s}\right) \\
		&-\ln \left(-s-\left(m_{2}^{2}-m_{1}^{2}\right)+2 q_{cmi}(s) \sqrt{s}\right) \\
		&\left.\left.-\ln \left(-s+\left(m_{2}^{2}-m_{1}^{2}\right)+2 q_{cmi}(s) \sqrt{s}\right)\right]\right\},
	\end{aligned}
	\label{eq:DR}
\end{equation}
$m_{1}$ and $m_{2}$ are the masses of the mesons in the loop, $\sqrt{s}$ is the energy of the system, $\mu$ is the regularization scale and $a_{ii}(\mu)$ is the subtraction constant. Besides, $q_{cmi}(s)$ is the three-momentum of the particle in the center-of-mass frame,
\begin{equation}
	\begin{aligned}
		q_{cmi}(s)=\frac{\lambda^{1 / 2}\left(s, m_{1}^{2}, m_{2}^{2}\right)}{2 \sqrt{s}}
	\end{aligned}
	\label{eq:qcmi}
\end{equation}
with the usual K\"allen triangle function $\lambda(a, b, c)=a^{2}+b^{2}+c^{2}-2(a b+a c+b c)$.

Since the dynamically generated states will be searched on the complex Riemann sheets, we need to extrapolate the loop function $G_{ii}(s)$ to the second Riemann sheet, i.e., 
\begin{equation}
	\begin{aligned}
		G_{ii}^{(I I)}(s)=&G_{ii}(s)-2 i \operatorname{Im} G_{ii}(s) \\
		=&G_{ii}(s)+\frac{i}{4\pi} \frac{q_{cmi}(s)}{\sqrt{s}}.
	\end{aligned}
	\label{eq:GII}
\end{equation}
The scattering amplitudes close to the pole can be written as \cite{Oller:2004xm, Guo:2006fu}
\begin{equation}
	\begin{aligned}
		T_{i j}(s)=\frac{g_{i} g_{j}}{s-s_{p}},
	\end{aligned}
	\label{eq:gij}
\end{equation}
where $g_i$ and $g_j$ are the couplings of the $i$-th and $j$-th channels, and $s_{p}$ is the square of the energy corresponding to the pole on the complex energy plane.

\section{Results}
\label{sec:Results}

Firstly, we determine the values of the regularization scale $\mu$ and the subtraction constant $a_{ii}(\mu)$ in the loop functions. In the studies of the $\bar{b}cq\bar{q}$ and $bc\bar{s}\bar{q}$ systems in Refs. \cite{Sakai:2017avl, Liu:2023hrz}, another regularization scheme of the three-momentum cutoff approach is applied to solve the singular integral, where the free parameter of the cutoff $q_{max}$ is taken from $400$ MeV to $600$ MeV. Due to the absence of the experimental data, following Refs. \cite{Oset:2001cn, Wu:2010jy, Dong:2021juy}, we take $\mu=q_{max}$, and match the real values of loop function from the two renormalization schemes at the threshold to determine the subtraction constant $a_{ii}(\mu)$. This ensures that the two methods can get similar results near the threshold and avoids the influence of singularity above the threshold of loop function with the three-momentum cutoff approach.

\begin{table}[htbp]
	\centering
	\setlength{\tabcolsep}{2mm}{
		\caption{The Masses (in MeV) of pseudoscalar and vector mesons and the thresholds (in MeV) of relevant channels.}
		\resizebox{1.0\textwidth}{!}
		{\begin{tabular}{lcccccccc}
				\hline\hline 
				States&$\rho$&$\omega$&$\bar{K}^{*}$&$\phi$&$\bar{D}$&$\bar{D}_{s}$&$\bar{D}^{*}$&$\bar{D}_{s}^{*}$\\ 
				Masses&$775.16$&$782.66$&$893.61$&$1019.46$&$1867.25$&$1968.35$&$2008.56$&$2112.20$\\ 
				States&$\bar{B}$&$\bar{B}^{*}$&$\bar{B}_{s}$&$\bar{B}_{s}^{*}$&&&&\\ 
				Masses&$5279.50$&$5324.71$&$5366.92$&$5415.40$&&&&\\ 
				\hline
				Channels&$\bar{B}\bar{D}$&$\bar{B}_{s}\bar{D}_{s}$&$\bar{B}^{*}\bar{D}$&$\bar{B}\bar{D}^{*}$&$\bar{B}_{s}^{*}\bar{D}_{s}$&$\bar{B}_{s}\bar{D}_{s}^{*}$&$\bar{B}^{*}\bar{D}^{*}$&$\bar{B}_{s}^{*}\bar{D}_{s}^{*}$\\ 
				Thresholds&$7146.75$&$7335.27$&$7191.96$&$7288.06$&$7383.75$&$7479.12$&$7333.27$&$7527.60$\\ 
				\hline\hline  
		\end{tabular}}
		\label{tab:Mass}}
\end{table}

\begin{figure}[htbp]
	\centering
	\includegraphics[width=0.5\linewidth,trim=0 0 0 0,clip]{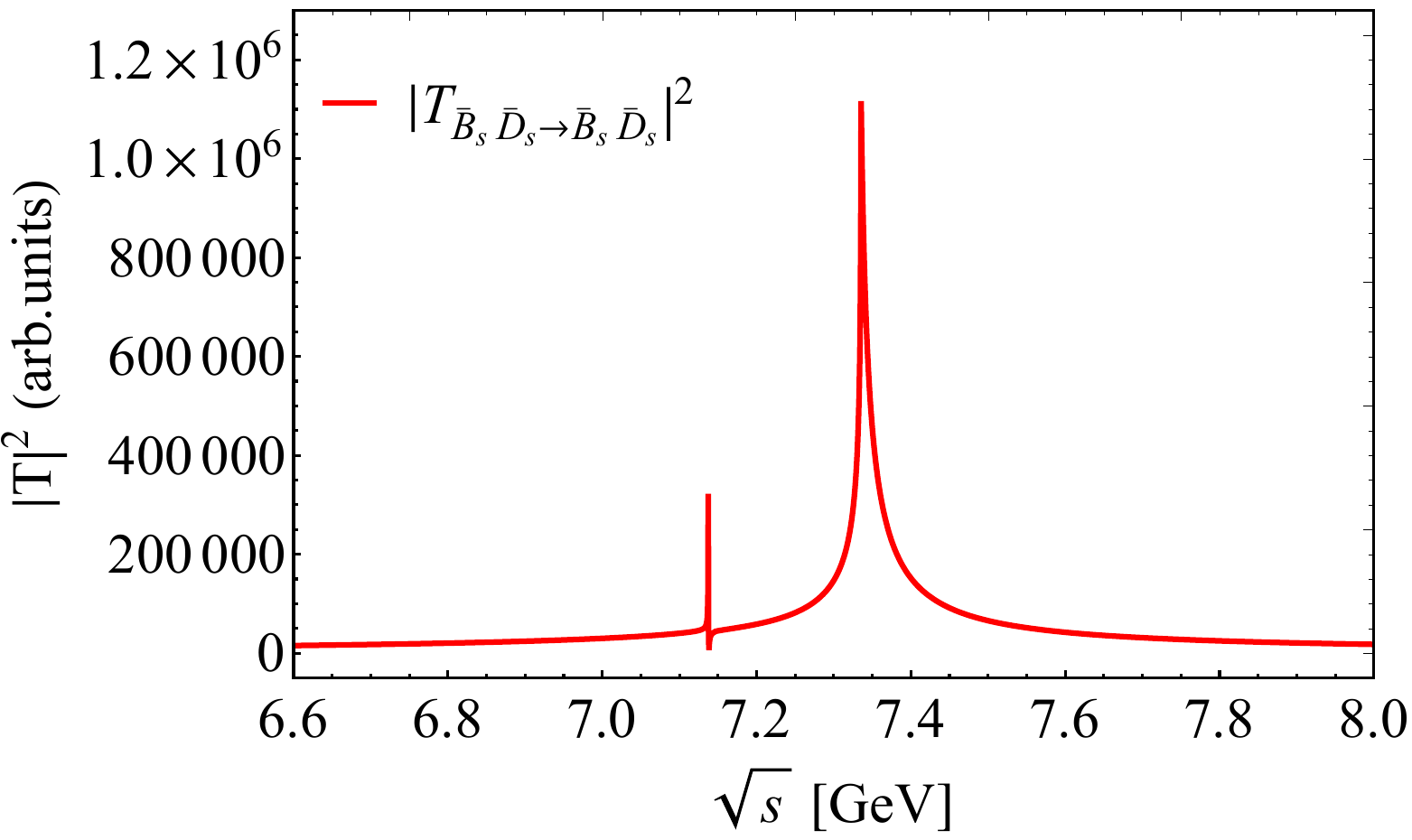} 
	\caption{The modulus square of $\bar{B}_{s}\bar{D}_{s}\to\bar{B}_{s}\bar{D}_{s}$ amplitude.}
	\label{fig:FigureJ011}
\end{figure}

\begin{figure}[htbp]
	\begin{subfigure}{0.45\textwidth}
		\centering
		\includegraphics[width=1\linewidth,trim=0 0 0 0,clip]{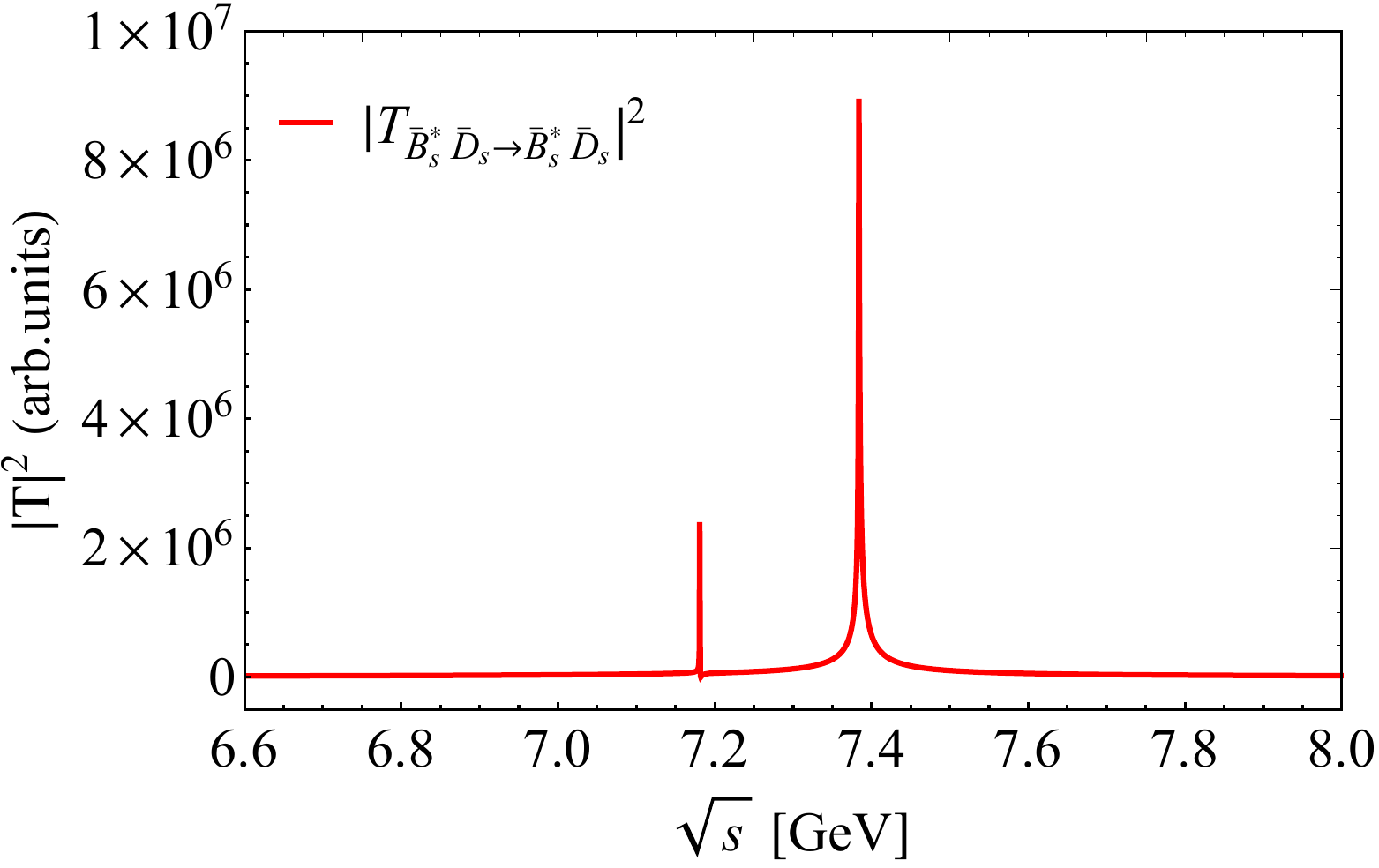} 
		\caption{\footnotesize }
		\label{fig:FigureJ111}
	\end{subfigure}
	\quad
	\quad
	\begin{subfigure}{0.45\textwidth}  
		\centering 
		\includegraphics[width=1\linewidth,trim=0 0 0 0,clip]{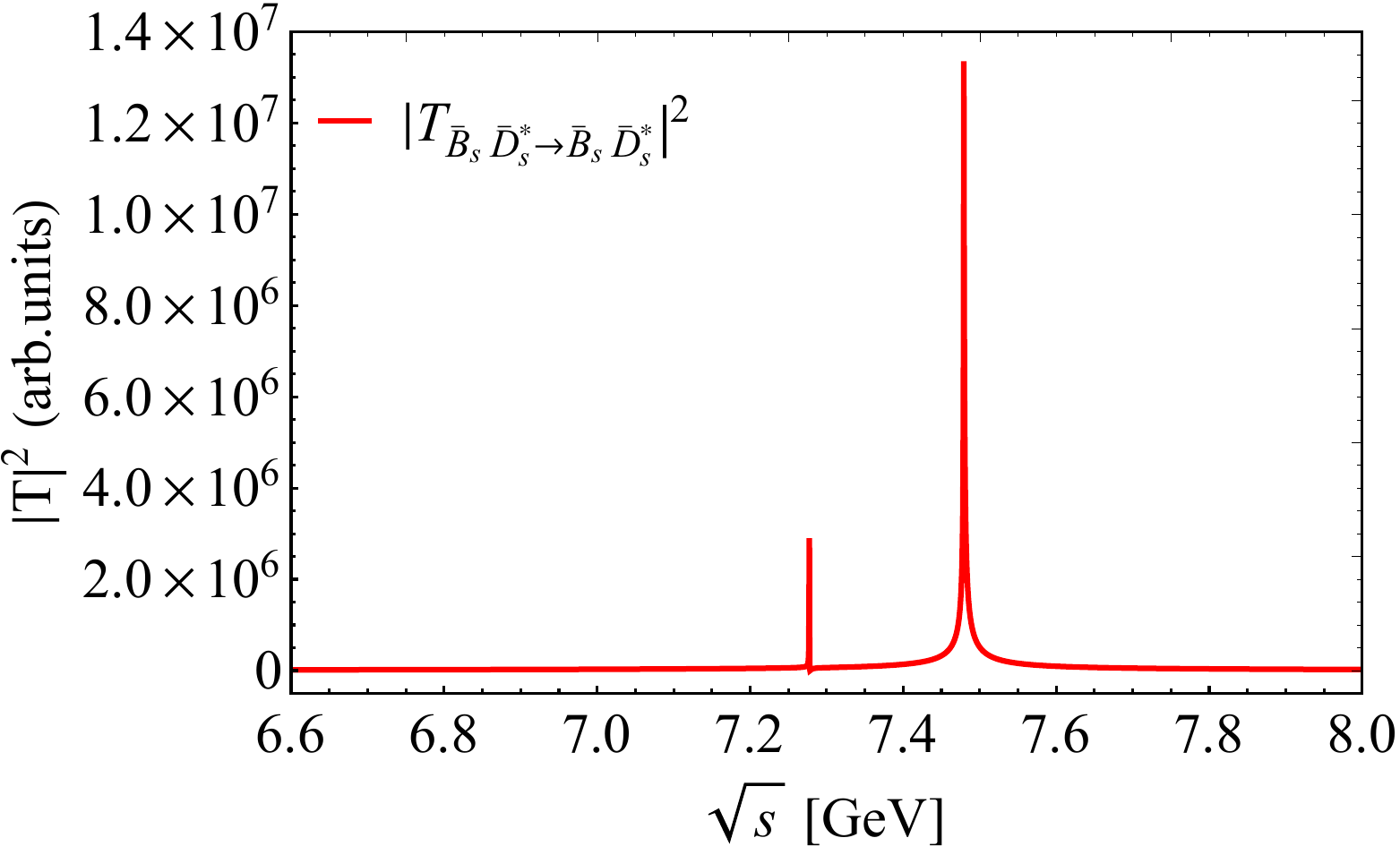} 
		\caption{\footnotesize }
		\label{fig:FigureJ112}  
	\end{subfigure}	
	\caption{The modulus square of $\bar{B}_{s}^{*}\bar{D}_{s}\to\bar{B}_{s}^{*}\bar{D}_{s}$ and $\bar{B}_{s}\bar{D}_{s}^{*}\to\bar{B}_{s}\bar{D}_{s}^{*}$ amplitudes.}
	\label{fig:FigureJ1}
\end{figure}

\begin{figure}[htbp]
	\begin{subfigure}{0.45\textwidth}
		\centering
		\includegraphics[width=1\linewidth,trim=0 0 0 0,clip]{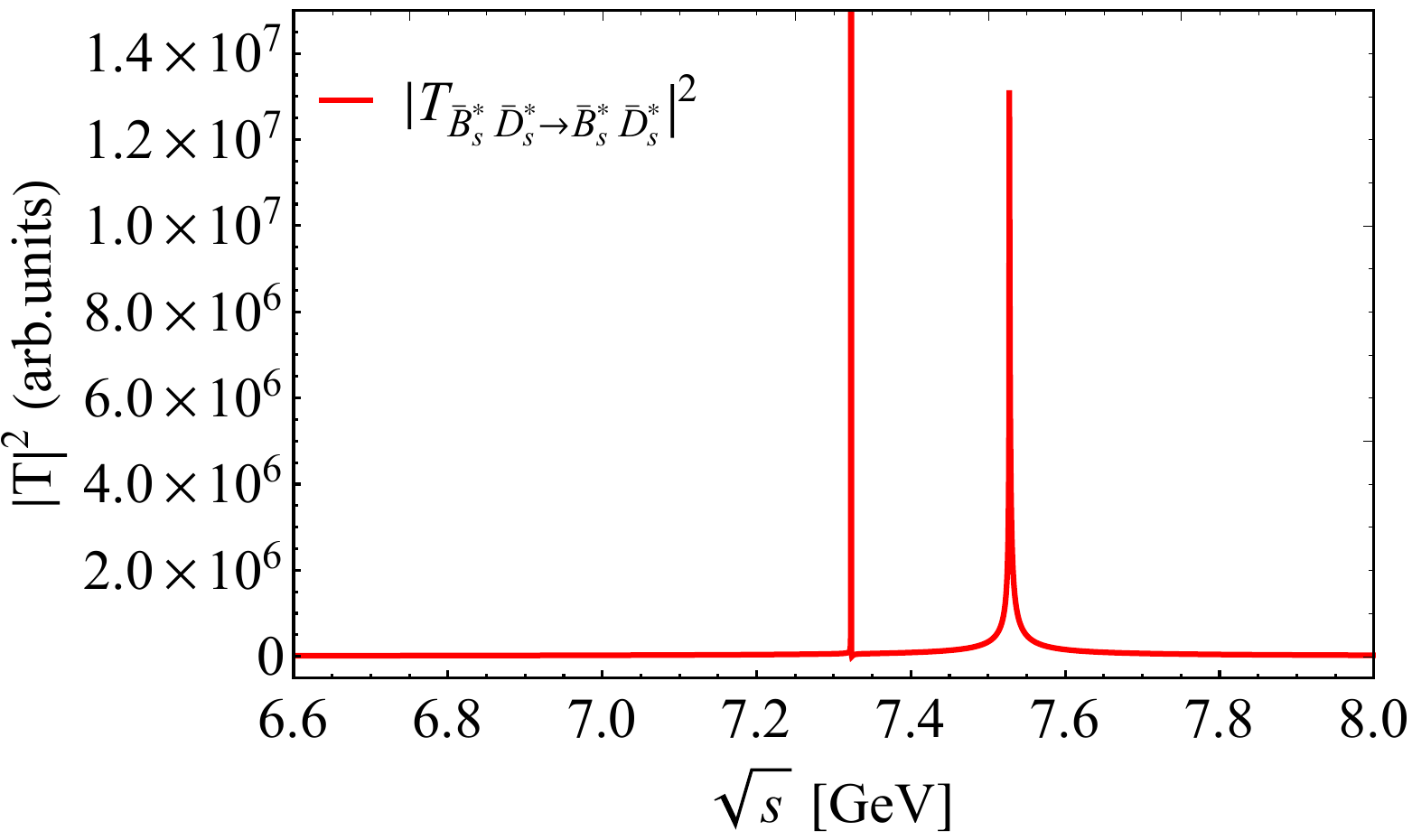} 
		\caption{\footnotesize}
		\label{fig:FigureVJ0}
	\end{subfigure}
	\quad
	\quad
	\begin{subfigure}{0.45\textwidth}  
		\centering 
		\includegraphics[width=1\linewidth,trim=0 0 0 0,clip]{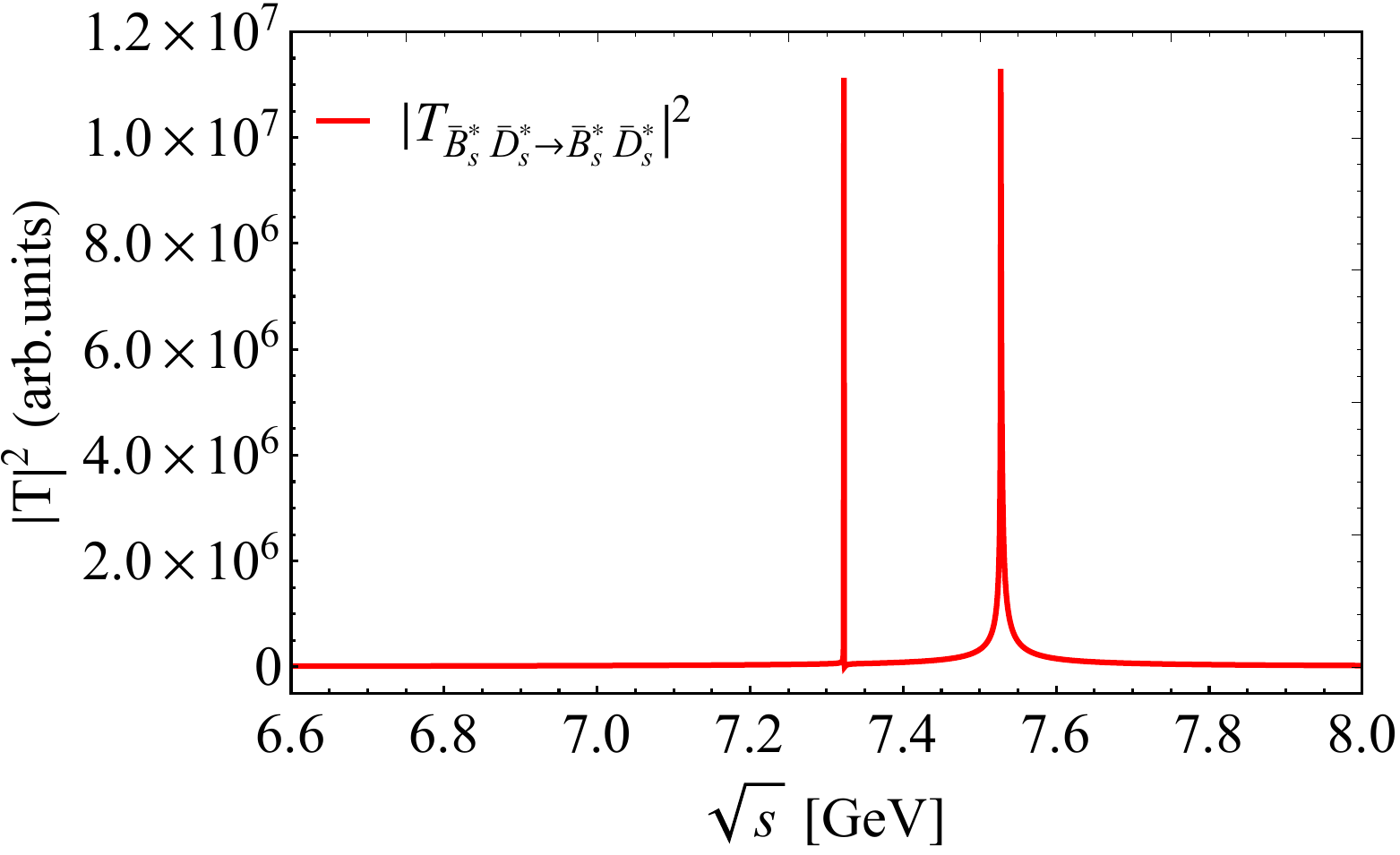} 
		\caption{\footnotesize}
		\label{fig:FigureVJ1}  
	\end{subfigure}	
	\begin{subfigure}{0.45\textwidth}  
		\centering 
		\includegraphics[width=1\linewidth,trim=0 0 0 0,clip]{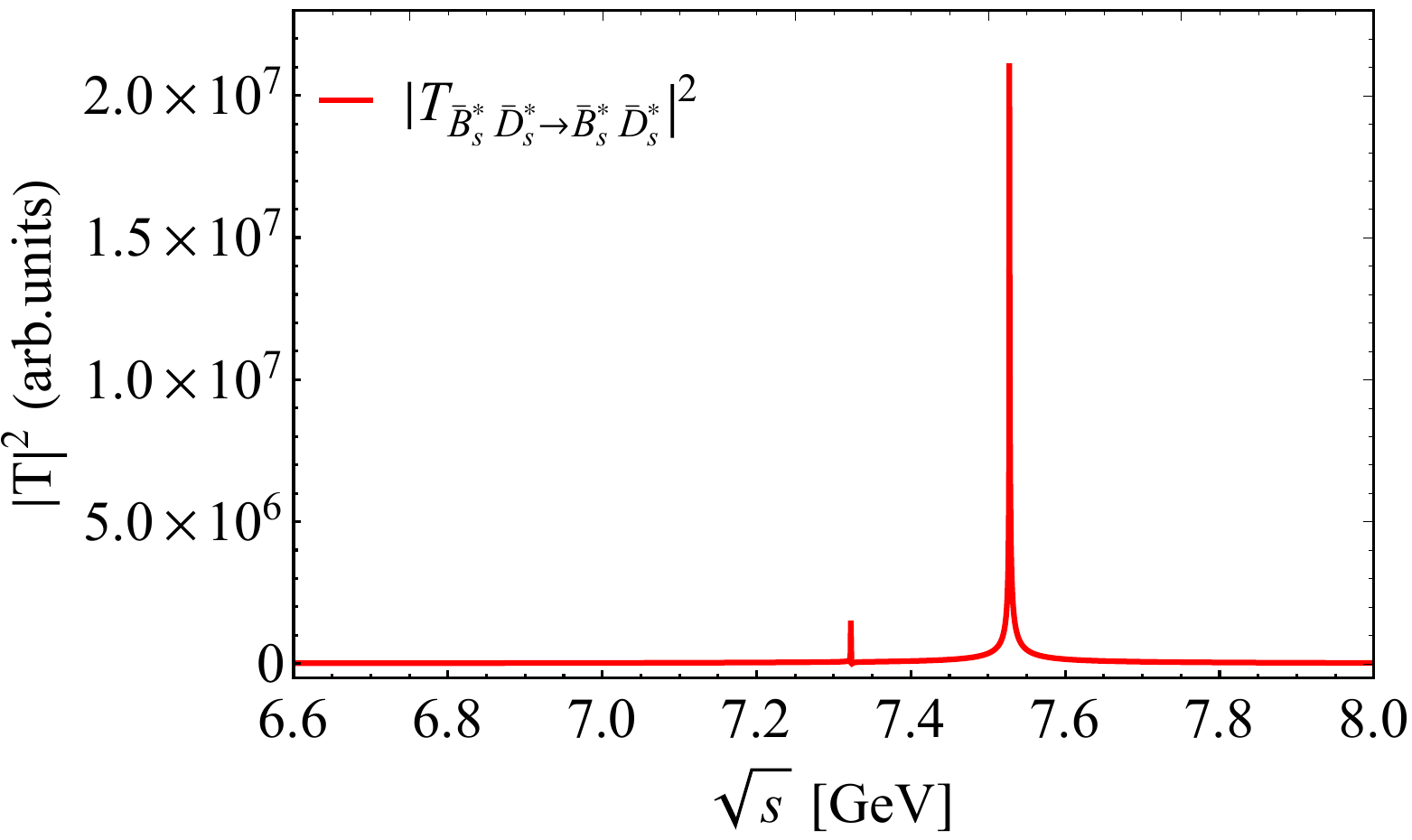} 
		\caption{\footnotesize}
		\label{fig:FigureVJ2}  
	\end{subfigure}	
	\caption{The modulus square of $\bar{B}_{s}^{*}\bar{D}_{s}^{*}\to\bar{B}_{s}^{*}\bar{D}_{s}^{*}$ amplitudes. (a) The total spin $J=0$. (b) The total spin $J=1$. (c) The total spin $J=2$.}
	\label{fig:FigureVJ}
\end{figure}

In Table \ref{tab:Mass}, we show the masses of the particles and the thresholds of the different channels needed in the calculations. 
At first we use $\mu=400$ MeV to perform numerical analyses. 
For the $\bar{B}\bar{D}$ and $\bar{B}_{s}\bar{D}_{s}$ coupled channels with the total spin $J=0$, we show the result of modulus square of $T_{\bar{B}_{s}\bar{D}_{s}\to\bar{B}_{s}\bar{D}_{s}}$ amplitude in Fig. \ref{fig:FigureJ011}, where two extremely narrow peaks locate near the thresholds. The first peak is about $10$ MeV below the $\bar{B}_{s}\bar{D}_{s}$ threshold and has a very small width, see Table \ref{tab:J0Coupling}. This state mainly couples to $\bar{B}_{s}\bar{D}_{s}$ indicating that it is a virtual state of $\bar{B}_{s}\bar{D}_{s}$, since the pole of this state locate on the Riemann sheet $(+-)$ where the signs of $-$ and $+$ represent the corresponding channel is open or close. In this case, the modulus square of amplitudes exhibit the cusp effect near threshold in all $\bar{B}_{s}^{(*)}\bar{D}_{s}^{(*)}$ systems, see Figs. \ref{fig:FigureJ1} and \ref{fig:FigureVJ}. The second one near the $\bar{B}\bar{D}$ threshold is the bound state of $\bar{B}\bar{D}$, which has been analyzed in details in Ref. \cite{Sakai:2017avl}. 

\begin{table}[htbp]
	\centering
	\setlength{\tabcolsep}{10mm}{
		\caption{The pole (in MeV) and its couplings (in GeV) to every channel in the two pseudoscalar mesons system. We use a minus sign of $-$ and a plus sign of $+$ to represent the corresponding channel is open or not in the parentheses after the pole position, and highlight the largest coupling of each pole in bold font.}
		\resizebox{1.0\textwidth}{!}
		{\begin{tabular}{lccc}
				\hline\hline 
				 $I(J^{P})$&Pole position&$|g_{\bar{B}\bar{D}}|$&$|g_{\bar{B}_{s}\bar{D}_{s}}|$ \\
				\hline 
				$0(0^{+})$&$7325.34-0.33i$ $(+-)$&$1.12$&$\bf{17.48}$ \\ 
				\hline\hline  
		\end{tabular}}
		\label{tab:J0Coupling}}
\end{table}

In Fig. \ref{fig:FigureJ1} and Table \ref{tab:J1Coupling}, the results of the pseudoscalar-vector system of spin $J=1$ are given. The two states located at $7382.82-0.25i$ MeV and $7478.57-0.18i$ MeV on the complex plane couple mostly to the $\bar{B}_{s}^{*}\bar{D}_{s}$ and $\bar{B}_{s}\bar{D}_{s}^{*}$ channels, respectively. And the couplings are $|g_i|=9.96$ GeV and $|g_i|=8.79$ GeV. This indicates that the first and second states are composed mainly of $\bar{B}_{s}^{*}\bar{D}_{s}$ and $\bar{B}_{s}\bar{D}_{s}^{*}$, respectively.

\begin{table}[htbp]
	\centering
	\setlength{\tabcolsep}{4mm}{
		\caption{The poles (in MeV) and their couplings (in GeV) to every channel in the pseudoscalar and vector mesons system.}
		\resizebox{1.0\textwidth}{!}
		{\begin{tabular}{lccccc}
				\hline\hline 
				$I(J^{P})$&Pole position&$|g_{\bar{B}^{*}\bar{D}}|$&$|g_{\bar{B}\bar{D}^{*}}|$&$|g_{\bar{B}_{s}^{*}\bar{D}_{s}}|$&$|g_{\bar{B}_{s}\bar{D}_{s}^{*}}|$ \\
				\hline 
				$0(1^{+})$&$7382.82-0.25i$ $(++-+)$&$0.97$&$0.00$&$\bf{9.96}$&$0.00$ \\ 
				$0(1^{+})$&$7478.57-0.18i$ $(+++-)$&$0.00$&$0.83$&$0.00$&$\bf{8.79}$ \\
				\hline\hline  
		\end{tabular}}
		\label{tab:J1Coupling}}
\end{table}

For the systems formed by the two vector mesons, we get three states $7526.96-0.17i$ MeV, $7526.83-0.18i$ MeV, and $7527.24-0.14i$ MeV on the complex Riemann sheet with the spin $J=0$, $1$, and $2$, respectively. And they all couple mostly to $\bar{B}_{s}^{*}\bar{D}_{s}^{*}$ channel. The results are shown in Fig. \ref{fig:FigureVJ} and Table \ref{tab:VJCoupling}. As discussed in Ref. \cite{Gamermann:2011mq}, these poles are not on the physical sheet while below the corresponding channel thresholds, indicating that they are virtual states. Besides, it is worth mentioning that the differences between the three poles come from the contributions of the contact terms, since the meson exchange processes have the same contributions.

\begin{table}[htbp]
	\centering
	\setlength{\tabcolsep}{10mm}{
		\caption{The poles (in MeV) and their couplings (in GeV) to every channel in the two vector mesons system.}
		\resizebox{1.0\textwidth}{!}
		{\begin{tabular}{lccc}
				\hline\hline 
				$I(J^{P})$&Pole position&$|g_{\bar{B}^{*}\bar{D}^{*}}|$&$|g_{\bar{B}_{s}^{*}\bar{D}_{s}^{*}}|$\\
				\hline
				$0(0^{+})$&$7526.96-0.17i$ $(+-)$&$0.80$&$\bf{9.13}$\\  
      			$0(1^{+})$&$7526.83-0.18i$ $(+-)$&$0.82$&$\bf{9.53}$\\  
				$0(2^{+})$&$7527.24-0.14i$ $(+-)$&$0.72$&$\bf{7.96}$\\ 
				\hline\hline  
		\end{tabular}}
		\label{tab:VJCoupling}}
\end{table}

Next, we take the regularization scale $\mu=600$ MeV.  
Then we find the corresponding poles on the different Riemann sheets, which are listed in Table \ref{tab:Mu600MeV}. 
Except for the first state generated by the pseudoscalar-vector system is the virtual state, the other ones are the quasi-bound states. 
Nevertheless, they have similar masses and widths with the virtual states in the case of $\mu=400$ MeV.
For the results in different parameters, we prefer that these systems can form the virtual states rather than quasi-bound states based on the following reason. 
With the chiral unitary approach, the experimental data of the heavy flavor state $T_{cc}$ can be well reconstructed by taking $q_{max}=415$ MeV in Ref. \cite{Feijoo:2021ppq}, while one needs $q_{max}=600$ MeV in the study of the low lying scalar mesons \cite{Liang:2014tia, Molina:2019udw, Ahmed:2020qkv, Wang:2021kka}. 

\begin{table}[htbp]
	\centering
	\setlength{\tabcolsep}{4mm}{
		\caption{The poles (in MeV) and their couplings (in GeV) to every channel in the case of $\mu=600$ MeV.}
		\resizebox{1.0\textwidth}{!}
		{\begin{tabular}{lccccc}
				\hline\hline 
				$I(J^{P})$&Pole position&&&&\\
				\hline
				&&$|g_{\bar{B}\bar{D}}|$&$|g_{\bar{B}_{s}\bar{D}_{s}}|$ &&\\ 
				$0(0^{+})$&$7334.49-0.09i$ $(+-)$&$0.56$&$\bf{9.40}$&& \\  
				\hline 
				&&$|g_{\bar{B}^{*}\bar{D}}|$&$|g_{\bar{B}\bar{D}^{*}}|$&$|g_{\bar{B}_{s}^{*}\bar{D}_{s}}|$&$|g_{\bar{B}_{s}\bar{D}_{s}^{*}}|$ \\
				$0(1^{+})$&$7381.98-0.33i$ $(--++)$&$1.11$&$0.00$&$\bf{11.79}$&$0.00$ \\ 
				$0(1^{+})$&$7476.84-0.35i$ $(---+)$&$0.00$&$1.15$&$0.00$&$\bf{12.56}$ \\
				\hline 
				&&$|g_{\bar{B}^{*}\bar{D}^{*}}|$&$|g_{\bar{B}_{s}^{*}\bar{D}_{s}^{*}}|$&&\\
				$0(0^{+})$&$7525.49-0.29i$ $(-+)$&$1.05$&$\bf{12.39}$&&\\  
				$0(1^{+})$&$7525.70-0.27i$ $(-+)$&$1.01$&$\bf{12.06}$&&\\  
				$0(2^{+})$&$7524.85-0.37i$ $(-+)$&$1.17$&$\bf{13.25}$&&\\ 
				\hline\hline  
		\end{tabular}}
		\label{tab:Mu600MeV}}
\end{table}

\section{Conclusions}
\label{sec:Conclusions}

We make a study of the $B_{c}$-like hadronic molecular states with the quark content $b\bar{c}s\bar{s}$ in this work. Three kinds of systems composed of the pseudoscalar-pseudoscalar, pseudoscalar-vector, and vector-vector mesons are calculated. The $S$-wave interactions are evaluated from the local hidden gauge Lagrangians which are extended to $SU(5)$ case, and the total scattering amplitudes are obtained by solving the Bethe-Salpeter equation. 

In the case of $\bar{B}\bar{D}$ and $\bar{B}_{s}\bar{D}_{s}$ coupled system, we get a state with the quantum number $I(J^{P})=0(0^{+})$ near the $\bar{B}_{s}\bar{D}_{s}$ threshold, which mainly couples to $\bar{B}_{s}\bar{D}_{s}$ channel. In the case of $\bar{B}^{*}\bar{D}$/$\bar{B}\bar{D}^{*}$/$\bar{B}_{s}^{*}\bar{D}_{s}$/$\bar{B}_{s}\bar{D}_{s}^{*}$ system, two states mainly composed of $\bar{B}_{s}^{*}\bar{D}_{s}$ and $\bar{B}_{s}\bar{D}_{s}^{*}$ are obtained, whose quantum numbers are $I(J^{P})=0(1^{+})$. For $\bar{B}^{*}\bar{D}^{*}$/$\bar{B}_{s}^{*}\bar{D}_{s}^{*}$ system, three states with $I(J^{P})=0(0^{+})$, $0(1^{+})$, and $0(2^{+})$ are found, which all mainly couple to the $\bar{B}_{s}^{*}\bar{D}_{s}^{*}$ channel. The slight differences of the masses and widths of these three states come from the contact terms contributions. Note that all the states above are of small widths and their masses are slightly below the corresponding thresholds. 
Under different parameter conditions in the pseudoscalar-vector and vector-vector systems, the masses and widths of the found states are similar, except for one case ($\mu$=400 MeV) where they are the virtual states and another case ($\mu$=600 MeV) where they are the quasi-bound states. We expect the experiments can search for these predicted hadronic molecules in the future.

\section*{Acknowledgements}

We would like to thank Prof. Xiang Liu, Fu-Lai Wang, and Si-Qiang Luo for valuable discussions. This work is supported by the National Natural Science Foundation of China under Grant No. 12247101. Zhi-Feng Sun thanks the support of the Fundamental Research Funds for the Central Universities under Grant No. lzujbky-2022-sp02 and the National Natural Science Foundation of China (NSFC) under Grants No. 11965016, 11705069 and 12047501.

 \addcontentsline{toc}{section}{References}
 
\end{document}